\documentclass[12pt]{article}
\usepackage{epsfig}
\usepackage{psfrag}
\usepackage{latexsym}
\def\beq{\begin{equation}}
\def\eeq{\end{equation}}
\def\bea{\begin{eqnarray}}
\def\eea{\end{eqnarray}}
\def\bet{\begin{tabular}}
\def\eet{\end{tabular}}
\def\bes{\begin{subequations}\bea}
\def\ees{\eea\end{subequations}}

\def\be{\begin{equation}}
\def\ee{\end{equation}}
\def\bc{\begin{center}}
\def\ec{\end{center}}
\def\bea{\begin{eqnarray}}
\def\eea{\end{eqnarray}}
\def\dd{\displaystyle}
\def\nn{\nonumber}

\catcode`@=11
\def\marginnote#1{}
\newcount\hour
\newcount\minute
\newtoks\amorpm
\hour=\time\divide\hour by60
\minute=\time{\multiply\hour by60 \global\advance\minute by-\hour}
\edef\standardtime{{\ifnum\hour<12 \global\amorpm={am}%
        \else\global\amorpm={pm}\advance\hour by-12 \fi
        \ifnum\hour=0 \hour=12 \fi
        \number\hour:\ifnum\minute<10 0\fi\number\minute\the\amorpm}}
\edef\militarytime{\number\hour:\ifnum\minute<10 0\fi\number\minute}
\def\draftlabel#1{{\@bsphack\if@filesw {\let\thepage\relax
   \xdef\@gtempa{\write\@auxout{\string
      \newlabel{#1}{{\@currentlabel}{\thepage}}}}}\@gtempa
   \if@nobreak \ifvmode\nobreak\fi\fi\fi\@esphack}
        \gdef\@eqnlabel{#1}}
\def\@eqnlabel{}
\def\@vacuum{}
\def\draftmarginnote#1{\marginpar{\raggedright\scriptsize\tt#1}}
\def\draft{\oddsidemargin 0.0truein
        \def\@oddfoot{\sl preliminary draft \hfil
        \rm\thepage\hfil\sl\today\quad\militarytime}
        \let\@evenfoot\@oddfoot \overfullrule 3pt
        \let\label=\draftlabel
        \let\marginnote=\draftmarginnote
   \def\@eqnnum{(\theequation)\rlap{\kern\marginparsep\tt\@eqnlabel}%
\global\let\@eqnlabel\@vacuum}  }
\catcode`@=12
\begin{document}
\begin{titlepage}
\vspace*{-1cm}
\phantom{hep-ph/******} 
\hfill{DFPD-07/TH/20}
\hfill{RM3-TH/07-18}
\hfill{CERN-PH-TH/2007-224}

\vskip 2.5cm
\begin{center}
\mathversion{bold}
{\Large\bf A SUSY SU(5) Grand Unified Model of Tri-Bimaximal Mixing from $A_4$}
\mathversion{normal}
\end{center}
\vskip 0.2  cm
\vskip 0.5  cm
\begin{center}
{\large Guido Altarelli}~\footnote{e-mail address: guido.altarelli@cern.ch}
\\
\vskip .1cm
Dipartimento di Fisica `E.~Amaldi', Universit\`a di Roma Tre
\\ 
INFN, Sezione di Roma Tre, I-00146 Rome, Italy
\\
\vskip .1cm
and
\\
CERN, Department of Physics, Theory Division
\\ 
CH-1211 Geneva 23, Switzerland
\\
\vskip .2cm
{\large Ferruccio Feruglio}~\footnote{e-mail address: feruglio@pd.infn.it}
\\
\vskip .1cm
Dipartimento di Fisica `G.~Galilei', Universit\`a di Padova 
\\ 
INFN, Sezione di Padova, Via Marzolo~8, I-35131 Padua, Italy
\\
\vskip .2cm
{\large Claudia Hagedorn}~\footnote{e-mail address: hagedorn@mpi-hd.mpg.de},
\\
\vskip .1cm
Max-Planck-Institut f\"ur Kernphysik
\\ 
Postfach 10 39 80, 69029 Heidelberg, Germany
\end{center}
\vskip 0.7cm
\begin{abstract}
We discuss a grand unified model based on SUSY SU(5) in extra dimensions and on the flavour group $A_4$ $\times$ U(1) which, besides reproducing tri-bimaximal mixing for neutrinos with the accuracy required by the data, also leads to a natural description of the observed pattern of quark masses and mixings. 

\end{abstract}
\end{titlepage}
\setcounter{footnote}{0}
\vskip2truecm
%%%%%%%%%%%%%%%%%%%%%%%% 1.  INTRODUCTION   %%%%%%%%%%%%%%%%%%%%%%%%%%%%%%
%
\section{Introduction}

It is an experimental fact \cite{data} that within measurement errors  
the observed neutrino mixing matrix is compatible with  
the so called tri-bimaximal (TB) form, introduced by Harrison, Perkins  
and Scott (HPS) \cite{hps}. The best measured neutrino mixing angle $\theta_{12}$ is just about 1$\sigma$ below the HPS value $\tan^2{\theta_{12}}=1/2$, while the other two angles are well inside the 1$\sigma$ interval \cite{data}. In a series of papers \cite{ma1,ma2,OurTriBi,Mod,us3} it has been  
pointed out that a broken flavour symmetry based on the discrete  
group $A_4$ appears to be particularly suitable to reproduce
this specific mixing pattern as a first approximation.
Other  
solutions based on alternative discrete or  continuous flavour groups have  
also been considered \cite{continuous,others}, but the $A_4$  
models have a very economical   
and attractive structure, e.g. in terms of group representations and of field content. 
 In most of the models $A_4$ is accompanied by additional symmetries, either continuous like U(1) or discrete like $Z_N$, which are necessary to eliminate unwanted couplings, to ensure the needed vacuum alignment and to reproduce the observed mass hierarchies.  
In this way one can construct natural models where the corrections to TB mixing can be evaluated in a well defined expansion.

Recently much attention has been devoted to the question whether a model for HPS mixing in the neutrino sector can be  suitably extended to also successfully describe the observed pattern of quark mixings and masses and whether this more complete framework can be made compatible with (supersymmetric (SUSY)) SU(5) or SO(10) grand unification. Early attempts of extending models based on $A_4$ to quarks  \cite{ma1.5,Mod} and to construct grand unified versions \cite{maGUT}  so far have not been completely satisfactory, e.g. do not offer natural mechanisms for mass hierarchies and for the vacuum alignment. A direct extension of the $A_4$ model to quarks leads to the identity matrix for $V_{CKM}$ in the lowest approximation, which at first looks promising. But the corrections 
 to it turn out to be strongly constrained by the leptonic sector, because lepton mixings are nearly TB, and are proven to be too small to accommodate the observed quark mixing angles \cite{Mod}. Also, the quark classification adopted in these models is not compatible with $A_4$ commuting with SU(5) \footnote{In ref. \cite{KM} an $A_4$ model compatible with the Pati-Salam group SU(4)$\times$ SU(2)$_L \times$ SU(2)$_R$ has been presented.}. 
 Due to this, larger discrete groups are considered for the description of quarks  and for grand unified versions with approximate TB mixing in the lepton sector. A particularly appealing set of models is based on the discrete group $T'$, the double covering group of $A_4$ \cite{T'0}. In ref. \cite{T'} a viable description was obtained, i.e. in the leptonic sector the predictions of the $A_4$ model are reproduced, while the $T'$ symmetry plays an essential role for reproducing the pattern of quark mixing. But, again, the classification adopted in this model is not compatible with grand unification. Unified models based on the discrete groups $T'$ \cite{CM}, $S_4$ \cite{S4} and $\Delta(27)$  \cite{27} have been 
discussed. Several models using the smallest non-abelian symmetry 
$S_3$ (which is isomorphic to $D_3$) can also be found in the recent literature \cite{S3}.

In conclusion, the group $A_4$ is considered by most authors to be too
limited to also describe quarks and to lead to a grand unified
description. In the present work we show that this negative attitude
is not justified and that it is actually possible to construct a
viable model based on $A_4$ which leads to a grand
unified theory (GUT) of quarks and leptons with TB mixing
for leptons. At the same time our model offers an example of an
extra dimensional GUT in which a description of all fermion masses
and mixings is attempted. The model is natural, since most of the
small parameters in the observed pattern of masses and mixings as well
as the necessary vacuum alignment are  justified by the symmetries of
the model. For this, it is sufficient to enlarge the  $A_4$ flavour
symmetry by adding a U(1) of the Froggatt-Nielsen type and to suitably
modify and extend the classification under the flavour group so that
finally all fermions transform in an SU(5) compatible way. 
 In addition, a $Z_3$ symmetry must be assigned to the fields
of the model which is, however, flavour-independent. The
formulation of SU(5) in extra dimensions has the usual advantages of
avoiding large Higgs representations to break SU(5) and of solving the
doublet-triplet splitting problem. A further ingredient of the model is a U(1)$_R$
symmetry which contains the discrete $R$-parity as a
subgroup. A see-saw realization
in terms of an $A_4$ triplet of right-handed neutrinos $N$ ensures the
correct ratio of light neutrino masses with respect to the GUT
scale. In the present model extra dimensional effects directly
contribute to determine the flavour pattern, in that the two lightest
tenplets $T_1$ and $T_2$ are in the bulk (with a doubling $T_i$ and
$T'_i$, $i=1,2$ to ensure the correct zero mode spectrum), whereas the
pentaplets $F$ and $T_3$ are on the brane. The hierarchy of quark and
charged lepton masses and of quark mixings is determined by a
combination of extra dimensional suppression factors for the first two
generations and of the U(1) charges, while the neutrino mixing angles
derive from $A_4$. The choice of the transformation properties of the two
 Higgses $H_5$ and $H_{\bar{5}}$ is also crucial. They are chosen to transform 
as two different $A_4$ singlets
$1$ and $1'$. As a consequence, mass terms for the Higgs colour
triplets are  not directly allowed \footnote{Even after $A_4$ breaking they are
forbidden at all orders by the U(1)$_R$ symmetry.} and their masses are
introduced by orbifolding, \`{a} la Kawamura \cite{5DSU5}. Finally, in this model, proton
decay is dominated by gauge vector boson exchange giving rise to
dimension six operators. Given the relatively large theoretical
 uncertainties, the decay rate is within the present
experimental limits. 

The resulting model is shown to be directly compatible with approximate TB mixing for leptons 
as well as with a realistic pattern of fermion masses and of quark mixings in a SUSY SU(5) 
framework. 

%
%%%%%%%%%%%%%%%%%%%%%%%% 2.  SECTION: MODEL   %%%%%%%%%%%%%%%%%%%%%%%%%%%%%%
%
\section{The Model}

We consider a SUSY GUT based on SU(5) in 4+1 dimensions. Leaving aside extra dimensional effects for a moment, from the four-dimensional (4D) point of view matter fields are chiral supermultiplets transforming 
as $10$, $\bar{5}$ and $1$ under SU(5). Part of the flavour symmetry is related to the discrete group $A_4$, whose properties 
are summarized, for instance, in section 2 of ref. \cite{Mod}, whose conventions are adopted here.
The three $\bar{5}$ and the three singlets (corresponding to the right-handed neutrinos) are grouped into $A_4$ triplets $F$ and $N$, while
the tenplets $T_1$, $T_2$ and $T_3$ are assigned to $1''$, $1'$ and $1$ singlets of $A_4$, respectively (see table 1).
The Higgs chiral supermultiplets that break the electroweak symmetry are $H_5$ and $H_{\bar{5}}$,
transforming as $(5,1)$ and $(\bar{5},1')$ under SU(5)$\times A_4$. We also consider a set of flavon
supermultiplets, all invariant under SU(5), that break the $A_4$ symmetry: two triplets $\varphi_T$ and 
$\varphi_S$ and two singlets $\xi$ and $\tilde{\xi}$. The alignment of their vacuum expectation values (VEVs) along appropriate directions in
flavour space will be the source of TB lepton mixing. It is well-known that, for this to work,  
each triplet should mainly contribute to the mass generation of a specific sector. At the leading order and after
spontaneous $A_4$ breaking, $\varphi_S$, $\xi$ and $\tilde{\xi}$ should give mass to neutrinos only, while $\varphi_T$ 
gives mass to charged leptons and to down quarks. This separation
can be realized with the help of an additional spontaneously broken
$Z_3$ symmetry under which $N$, $F$, $T_i$, $H_{5,\bar{5}}$, $\varphi_S$, $\xi$ and $\tilde{\xi}$ are multiplied by $\omega=\exp(i2\pi/3)$, while 
$\varphi_T$ is left invariant. The generation of the up quark masses as well as the 
quark mixings will be discussed below.

The breaking of the grand unified symmetry is a potential source of serious problems, like those related to the doublet-triplet splitting
and to proton decay. One of the most efficient mechanisms to break SU(5) and avoid these problems is the one based on compactification of extra spatial dimensions \cite{5DSU5}.
The simplest setting is an SU(5) gauge invariant five-dimensional (5D) theory where the fifth dimension is compactified on a circle $S^1$ of radius $R$.
The gauge fields, living in the whole 5D space-time, are assumed to be periodic along the extra dimension only up to
a discrete parity transformation $\Omega$ such that the gauge fields of the SU(3)$\times$SU(2)$\times$U(1) subgroup are periodic, 
while those of the coset SU(5)/SU(3)$\times$SU(2)$\times$U(1) are antiperiodic. Only the gauge vector bosons of SU(3)$\times$SU(2)$\times$U(1)
possess a zero mode. Those of SU(5)/SU(3)$\times$SU(2)$\times$U(1)
form a Kaluza-Klein tower starting at the mass level
$1/R$. From the viewpoint of a 4D observer, these boundary conditions effectively break SU(5) down to the Standard Model (SM) gauge group,
at a GUT scale of order $1/R$. The transformation $\Omega$ is an automorphism of the SU(5) algebra, so that the whole construction 
can be carried out within an SU(5) invariant formalism. 
An important advantage of this mechanism is that it provides a simple solution to the doublet-triplet splitting problem. 
The parity $\Omega$ is consistently extended to the Higgs multiplets $H_5$ and $H_{\bar 5}$, also assumed to live in the whole 5D space, in such a way that the electroweak doublets are periodic,
whereas the colour triplets are antiperiodic. In this way we have zero modes only for the doublets and the lightest colour triplets get masses of order $1/R$.
Notice that, if the model is supersymmetric as in the case under discussion here, we have an effective 4D $N=2$ SUSY, induced by the
original $N=1$ SUSY in five dimensions. To reduce $N=2$ down to $N=1$ it is convenient to compactify the fifth dimension on the orbifold 
$S^1/Z_2$ rather than on the circle $S^1$. The orbifold projection eliminates all the zero modes of the extra states belonging to $N=2$ SUSY
and also those of the fifth component of the gauge vector bosons. The zero modes we are left with are the 4D gauge bosons of the
SM, two electroweak doublets and their $N=1$ SUSY partners. To complete the solution of the doublet-triplet splitting problem, we should also forbid 
a large mass term $H_{5} H_{\bar 5}$, which would otherwise lift the doublet masses. 
 As will be explained below, this is automatically guaranteed by
the U(1)$_R$ symmetry that we specify in table 1. 

For the gauge vector bosons and the Higgses $H_{5}$ and $H_{\bar 5}$ we will adopt this setup, which is described in detail in refs. \cite{5D}.
For the remaining fields we have much more freedom \cite{5D,5Dfreedom}. Indeed the orbifold $S^1/Z_2$ corresponds to a segment where the fifth coordinate $y$ runs from 0 to $\pi R$.
The boundaries of the segment determine two 4D slices of the original 5D space-time. When boundary conditions are consistently defined for the local parameters
of SU(5) gauge transformations, we find that such transformations are generically non-vanishing only in the bulk and at $y=0$. At the opposite endpoint
of the segment, $y=\pi R$, the only gauge transformations that are different from zero are those of the SM. Therefore we have three qualitatively different possible
locations for the remaining fields: in the bulk, at the SU(5) preserving brane $y=0$, or at the SU(5) breaking brane $y=\pi R$. We choose to put the two tenplets $T_1$ and $T_2$ of the first and second family in the bulk. As explained in ref.  \cite{5D, 5Dfreedom} to obtain the correct zero mode spectrum with intrinsic parities compatible with symmetry and orbifolding, one must introduce two copies of each multiplet with opposite parity $\Omega$ in the bulk. 
Therefore $T_{1,2}$ is a short notation for the copies $T_{1,2}$ and $T_{1,2}'$. The zero modes of $T_{1,2}$ are the SU(2) quark doublets $Q_{1,2}$,
while those of $T_{1,2}'$ are $U_{1,2}^c$ and $E_{1,2}^c$. All remaining $N=1$ supermultiplets are assigned to the SU(5) preserving brane at $y=0$.
\\[0.2cm]
\begin{table}[h]
\begin{center}
\begin{tabular}{|c||c|c|c|c|c||c|c||c|c|c|c|c||c|c|c|}
\hline
{\tt Field}& $N$& $F$ & $T_1$ & $T_2$ & $T_3$ & $H_5$ & $H_{\bar 5} $ & 
$\varphi_T$ & $\varphi_S$ & $\xi,~\tilde{\xi}$ & $\theta$& $\theta''$  & $\varphi_0^T$ & $\varphi_0^S$ & $\xi_0$ \\
\hline
SU(5) & $1$& $\bar{5}$ & $10$ & $10$ & $10$ & $5$ & $\bar{5}$ & 
$1$ & $1$ & $1$ & $1$ & $1$ & $1$ & $1$ & $1$ \\
\hline
$A_4$ & $3$& $3$ & $1''$ & $1'$ & $1$ & $1$ & $1'$ & 
$3$ & $3$ & $1$ & $1$ & $1''$ & $3$ & $3$ & $1$ \\
\hline
U(1) & $0$& $0$ & $3$ & $1$ & $0$ & $0$ & $0$ &
$0$ & $0$ & $0$ & $-1$ & $-1$ & $0$ & $0$ & $0$ \\
\hline
$Z_3$ & $\omega$& $\omega$ & $\omega$ & $\omega$ & $\omega$ & $\omega$ & $\omega$ &
$1$ & $\omega$ & $\omega$ & $1$ & $1$ & $1$ & $\omega$ & $\omega$ \\
\hline 
U(1)$_R$ & $1$& $1$ & $1$ & $1$ & $1$ & $0$ & $0$ & 
$0$ & $0$ & $0$ & $0$ & $0$ & $2$ & $2$ & $2$ \\
\hline
\end{tabular}
\end{center}
\caption{Fields and their transformation properties under SU(5), $A_4$, U(1), $Z_3$ and U(1)$_R$. $T_1$ and $T_2$ come in two replicas with the same quantum numbers, except for the intrinsic parity $\Omega$. For simplicity, we only show one of them in the table.}
\end{table}
\vspace{0.2cm}

An interesting feature of the 5D setup is the automatic suppression of the Yukawa couplings for the fields living in the bulk.
Indeed, a bulk field $B$ and its zero mode $B^0$ are related by:
\be
B=\frac{1}{\sqrt{\pi R}} B^0+...
\ee
where dots stand for the higher modes. This expansion produces a suppression factor
\be
s\equiv\dd\frac{1}{\sqrt{\pi R \Lambda}}<1~~~.
\label{vsup}
\ee  
Thereby, $\Lambda$ denotes the ultraviolet cut-off. Such a suppression factor enters the Yukawa couplings depending on the field $B^0$. 
As a result, the hierarchies among the charged fermion masses are partly
due to the geometrical dilution of the Yukawa couplings involving $T_{1,2}$. However this dilution cannot account for all the observed hierarchies
and, to achieve a realistic mass spectrum, we also exploit the Froggatt-Nielsen mechanism.
The tenplets $T_1$ and $T_2$ are charged under a U(1) flavour group, spontaneously broken by the VEVs of two fields
$\theta$ and $\theta''$ both carrying U(1) charges $-1$. The elements of the charged fermion mass matrices are
provided by higher-dimensional operators with powers of $\theta$ and $\theta''$ balancing the U(1) charge of the relevant
combination of matter fields. Indeed, we need two fields, $\theta$ and $\theta''$, in order to reproduce a realistic pattern of
quark masses and mixing angles. Under $A_4$, $\theta$ is invariant, while $\theta''$ transforms as 
$1''$.
All this is summarized in table 1. 

Notice that, once we have introduced all the fields with the quantum numbers displayed
in table 1, there will be no contribution coming from colour triplet exchange to the dangerous dimension five operator that induces proton decay
in SUSY theories. Actually that operator is strictly forbidden as long as the U(1)$_R$ symmetry remains unbroken. 
Indeed, the superpotential  of the effective $N=1$ SUSY should have U(1)$_R$ charge $+2$, to compensate the $R$-charge $-2$ coming
from the Grassmann integration measure $d^2\theta$. With the $R$ assignment in table 1, all superpotential couplings bilinear in the matter fields
$N$, $F$ and $T$ have $R$-charge $+2$ and are allowed. At the same time dangerous operators are forbidden. First of all these include
the mass term $H_5 H_{\bar 5}$, that would spoil the solution to the doublet-triplet splitting problem. Moreover, since U(1)$_R$ contains
the discrete $R$-parity, also all renormalizable baryon and lepton number violating operators, such as $F H_5$ and $F F T$, are not allowed. 
Finally, the dimension five operator $FTTT$, leading to proton decay, has $R$-charge $+4$ and 
 therefore is absent. 
As discussed in detail in ref. \cite{Mod} and briefly recalled in section 4, the U(1)$_R$ symmetry plays also an important role in the dynamics that selects the correct vacuum of the
theory, which is a crucial feature to reproduce nearly TB mixing in the lepton sector. The U(1)$_R$ symmetry is a remnant of the SU(2)$_R$ symmetry of the
$N=2$ SUSY bulk action, before compactification. By reducing $N=2$ down to $N=1$ the orbifold projection breaks SU(2)$_R$ down to U(1)$_R$.  
Eventually, after the inclusion of $N=1$ SUSY breaking effects, the U(1)$_R$ symmetry will be broken down to the discrete $R$-parity, at the low energy scale $m_{SUSY}$.
The operator $FTTT$ might be generated, but with a highly suppressed coupling of the kind $(m_{SUSY}/\Lambda)^n/ \Lambda$, $n>0$.
Therefore, the leading contribution
to proton decay comes from gauge vector boson exchange and the corresponding proton decay rate is typically small enough, though suffering from considerable uncertainties \cite{5Dpdecay}.

%
%%%%%%%%%%%%%%%%%%%%%%%% 3. SECTION: FERMION MASSES   %%%%%%%%%%%%%%%%%%%%%%%%%%%%%%
%
\section{Fermion Masses}
The $N=2$ SUSY invariance is broken down to $N=1$ by the orbifold projection, but it still forbids 5D superpotential couplings. These couplings should be strictly
localized at one of the two branes. By choosing the brane at $y=0$, the brane action reads:
\be
\int d^4x\int_0^{\pi R} dy \int d^2\theta~ w(x) \delta(y)+h.c.=\int d^4x \int d^2\theta~ w(x) +h.c.~~~.
\ee
The superpotential $w$, which can be expressed in terms of $N=1$ superfields, can be decomposed into several parts:
\be
w=w_{up}+w_{down}+w_{\nu}+w_d+...~~~.
\label{superw}
\ee
The first three contributions in eq. (\ref{superw}) give rise to fermion masses
after $A_4$, U(1) and electroweak symmetry breaking. They are of the form:
\bea
w_{up}&=& 
\frac{1}{\Lambda^{1/2}} H_5 T_3 T_3+
\frac{{\theta''}}{\Lambda^2} H_5 T_2 T_3 +
\frac{{\theta''}^2}{\Lambda^{7/2}} H_5 T_2 T_2  +  \frac{\theta{\theta''}^2}{\Lambda^4} H_5 T_1 T_3 \nn\\
&+& \frac{{\theta}^4}{\Lambda^{11/2}} H_5 T_1 T_2 +
\frac{{\theta}{\theta''}^3}{\Lambda^{11/2}} H_5 T_1 T_2 +
\frac{{\theta}^{5}{\theta''}}{\Lambda^{15/2}} H_5 T_1 T_1
+ \frac{\theta^2{\theta''}^{4}}{\Lambda^{15/2}} H_5 T_1 T_1
\label{wup}
\eea
\bea
w_{down}&=&
\frac{1}{\Lambda^{3/2}} H_{\bar 5} (F \varphi_T)''  T_3+
\frac{\theta}{\Lambda^3} H_{\bar 5} (F \varphi_T)' T_2 +
\frac{\theta^3}{\Lambda^5} H_{\bar 5} (F \varphi_T)  T_1+\frac{{\theta''}^3}{\Lambda^5} H_{\bar 5} (F \varphi_T)  T_1 \nn\\
&+& 
\frac{{\theta''}}{\Lambda^3} H_{\bar 5} (F \varphi_T)''  T_2 +
\frac{\theta^2 {\theta''}}{\Lambda^5} H_{\bar 5} (F \varphi_T)'   T_1 +
\frac{\theta{\theta''}^2}{\Lambda^5} H_{\bar 5} (F \varphi_T)''  T_1 +...~~~,
\label{wdown}
\eea
where dots stand for higher-dimensional operators.
In both, $w_{up}$ and $w_{down}$, the dimensionless coefficients of each independent operator have been omitted, for notational simplicity.
They are not predicted by the flavour symmetry, though they are all expected to be of the same order.  
The powers of the cut-off $\Lambda$ are determined by the dimensionality of the various operators, by recalling that brane and bulk superfields have mass dimensions 1
and 3/2, respectively. Some combinations of matter fields, as for instance $T_1 T_2$ in $w_{up}$, appear several times, but with the same cut-off suppression. 
Provided $\theta$ and $\theta''$ develop VEVs of similar size, the corresponding contributions to the charged fermion mass matrices will be of the same order. 
The bulk matter supermultiplets $T_1$ and $T_2$ come in two copies and, to keep our notation compact, the previous formulae do not contain all possible terms originating from such a doubling. 
For instance, $F_1 T_2$ stands for both combinations $F_1 T_2$ and $F_1 T_2'$,  which are suppressed by the same 
power of $\Lambda$, but can differ by order-one relative weights.  It is important to keep this point in mind, since it allows to escape the too rigid mass
relations between the first two generations of charged leptons and down quarks  predicted by the minimal SU(5) GUT. 

Neutrinos have both Dirac and Majorana mass terms, induced by:
\be
w_{\nu}=\frac{y^{D}}{\Lambda^{1/2}} H_5 (N F)+(x_a \xi +\tilde{x}_a \tilde{\xi})(NN)+x_b (\varphi_S NN)~~~,
\label{wnu}
\ee
where $\tilde{\xi}$ is defined as the combination of the two independent $\xi$-type fields 
 which has a vanishing VEV. Therefore, it does not contribute to the 
neutrino masses. 

The last term in eq. (\ref{superw}), $w_d$, is responsible for the alignment of the flavon fields $\varphi_T$,
$\varphi_S$, $\xi$ and $\tilde{\xi}$. The fields $\theta$ and $\theta''$ get VEVs from the minimisation
of the D-term of the scalar potential. We will discuss these issues in the next section.
For the time being we assume that the scalar components of the supermultiplets acquire VEVs according to
the following scheme:
\bea
\dd\frac{\langle \varphi_T \rangle}{\Lambda}=(v_T,0,0)~~~,&
\dd\frac{\langle \varphi_S \rangle}{\Lambda}=(v_S,v_S,v_S)~~~,&
\dd\frac{\langle \xi \rangle}{\Lambda}=u~~~,\nn\\
\dd\frac{\langle \theta \rangle}{\Lambda}=t~~~,& 
\dd\frac{\langle \theta'' \rangle}{\Lambda}=t'' ~~~.&
\label{align}
\eea
The Higgs multiplets live in the bulk and what matters for the Yukawa couplings are the values of the VEVs at $y=0$:
\be
\langle H_5(0) \rangle=\dd \frac{v_u^0}{\sqrt{\pi R}}~~~,~~~~~~~~~~ \langle H_{\bar 5}(0) \rangle= \dd\frac{v_d^0}{\sqrt{\pi R}}~~~,
\label{hvev}
\ee
where $v_{u,d}^0$ have mass dimension 1. The electroweak scale is determined by the relation:
\be
v_u^2+v_d^2\approx (174~{\rm GeV})^2~~~,~~~~~v_u^2\equiv\int_0^{\pi R} dy \left\vert\langle H_5(y) \rangle\right\vert^2~~~,~~~~
v_d^2\equiv\int_0^{\pi R} dy \left\vert\langle H_{\bar 5}(y) \rangle\right\vert^2~~~.
\ee
Notice that the electroweak gauge boson masses depend on the 5D averages of $\vert\langle H_{5,\bar{5}}(y) \rangle\vert^2$, rather than on the values at $y=0$.
If the VEVs of $H_{5,\bar{5}}$ are constant along the fifth dimension, then $v_u^0=v_u$ and $v_d^0=v_d$.
However, if the profile of $\langle H_{5,\bar{5}}(y) \rangle$ is not flat in $y$, the parameters $v_{u,d}^0$ are less constrained. 
In order to obtain $v_{u,d}^0\ne v_{u,d}$, we need some special dynamics on the $y=0$ and $y=\pi R$ branes,
that we cannot control without detailing additional features of the model, such as the breaking of the residual $N=1$ SUSY
and the generation of a non-trivial potential for the electroweak doublets. In this section we consider  $v_{u,d}^0\ne v_{u,d}$ as an open possibility and we will discuss
a possible application of it.
All the other fields have vanishing VEVs.

From these VEVs, the superpotential terms in eqs. (\ref{wup},\ref{wdown},\ref{wnu}) and the volume suppression $s$ of eq. (\ref{vsup}), it is immediate to derive the 
fermion mass matrices. In the up and down quark sector we get, 
up to unknown coefficients of order one for each
matrix element and by adopting the convention $\overline{f_R} m_f f_L$:
\be
m_u=\left(
\begin{array}{ccc}
s^2 t^5 {t''}+s^2 t^2 {t''}^4
& s^2 t^4+s^2 t {t''}^3&
s t {t''}^2\\
s^2 t^4+s^2 t {t''}^3&s^2 {t''}^2& s t''\\
s t {t''}^2&s t'' & 1
\end{array}
\right) s v_u^0~~~,
\label{mu}
\ee
\be
m_d=\left(
\begin{array}{ccc}
s t^3+s {t''}^3 & ... & ... \\
s t^2{t''}&s t& ...\\
s t {t''}^2& s t''& 1
\end{array}
\right) v_T s v_d^0 ~~~,
\label{md}
\ee
where the dots stand for subleading contributions, that will be fully discussed in section 5. 
Here we explicitly see the interplay between the volume dilution and the Froggatt-Nielsen mechanism, to 
 achieve the hierarchical pattern
of the quark mass matrices. Realistic values of quark mass ratios and mixing angles are obtained by assuming
\be
t\approx t''\approx s\approx O(\lambda)~~~~\mbox{with}~~~~\lambda\equiv 0.22~~~.
\label{ttpr_order}
\ee
Indeed, with this choice we obtain
\be
m_u=\left(
\begin{array}{ccc}
\lambda^8& \lambda^6& \lambda^4\\
\lambda^6& \lambda^4& \lambda^2\\
\lambda^4& \lambda^2& 1
\end{array}
\right) \lambda v_u^0~~~,
\label{mulam}
\ee
\be
m_d=\left(
\begin{array}{ccc}
\lambda^4 & ... & ...\\
\lambda^4& \lambda^2& ...\\
\lambda^4& \lambda^2& 1
\end{array}
\right) v_T \lambda v_d^0 ~~~.
\label{mdlam}
\ee
We anticipate that, in the absence of corrections to the vacuum alignment given in eq. (\ref{align}), the dots receive contributions from highly suppressed operators. 
In this case the entries 12, 13 and 23 of $m_d/(v_T v_d^0)$ would be of order $\lambda^{7}$, $\lambda^5$ and $\lambda^5$, respectively.
Since $v_T\approx O(\lambda^2)$ (see below),  $m_b/m_t\approx v_T v_d^0/v_u^0 \approx \lambda^2$ is easily reproduced by taking
$v_u^0\approx v_d^0$.
Notice that there is an overall factor $s\approx O(\lambda)$, coming from the normalization of the Higgs VEVs,
eq. (\ref{hvev}), suppressing both $m_u$ and $m_d$. In order to avoid large dimensionless coefficients, we make use of the freedom related to
the boundary values $v_{u,d}^0$ and we will assume that $v_{u,d}\approx \lambda v_{u,d}^0$. In this way, the Yukawa coupling of the top quark
is of order one and, by the patterns given in eqs. (\ref{mulam},\ref{mdlam}), also all the other couplings are of the same order.
Alternatively, if the Higgs VEVs are flat along the fifth dimension and $v_{u,d}^0=v_{u,d}$, we must assume that all Yukawa operators in $w$ have similar 
couplings of order $1/\lambda$ \cite{Nomura}. 
To correctly reproduce the quark mixing angle between the first and the second generation, a moderate tuning is needed in order to enhance
the individual contributions from the up and down sectors, which are both of order $\lambda^2$.

The mass matrix for the charged lepton sector is of the type:
\be
m_e=\left(
\begin{array}{ccc}
s t^3+s {t''}^3 &s t^2{t''}&s t {t''}^2 \\
...&s t& s t'' \\
...&...& 1
\end{array}
\right) v_T s v_d^0
=\left(
\begin{array}{ccc}
\lambda^4 & \lambda^4& \lambda^4\\
...& \lambda^2&\lambda^2\\
...& ...& 1 
\end{array}
\right) v_T \lambda v_d^0 ~~~.
\label{me}
\ee
We observe that the minimal SU(5) relation $m_e=m_d^T$ is relaxed. Indeed, while the third column of $m_d$ exactly coincides with the
third row of $m_e$, thus implying $m_b\approx m_\tau$ at the GUT scale, the remaining entries are only equal (up to a transposition) at the level
of the orders of magnitude, since $T_{1,2}$ are doubled. This allows to evade the too rigid relations $m_\mu=m_s$ and $m_e=m_d$ of minimal SU(5). In our 5D setup these relations 
hold only up to order one coefficients and acceptable values of the masses for $e$, $\mu$, $d$ and $s$ can be accommodated.
 
In the neutrino sector, after the fields $\varphi_S$ and $\xi$ develop their VEVs, the gauge singlets $N$ become heavy and the see-saw mechanism
takes place. The mass matrix for light neutrinos is given by:
\be
m_\nu=
\frac{1}{3 a(a+b)}
\left(
\begin{array}{ccc}
3a+b& b& b\\
b& \dd\frac{2a b+b^2}{b-a}& \dd\frac{b^2-a b -3a^2}{b-a}\\
b& \dd\frac{b^2-a b-3 a^2}{b-a}& \dd\frac{2a b+b^2}{b-a}
\end{array}
\right) \frac{s^2 (v_u^0)^2}{\Lambda} ~~~,
\label{mnu}
\ee
where 
\be
a\equiv \frac{2 x_a u}{(y^{D})^2}~~~,~~~~~~~b\equiv \frac{2 x_b v_S}{(y^D)^2}~~~.
\label{ad}
\ee
The neutrino mass matrix is diagonalized by the transformation:
\be
U^T m_\nu U ={\tt diag}(m_1,m_2,m_3)~~~,
\ee
where, in units of $s^2 (v_u^0)^2/\Lambda$,
\be
m_1=\dd\frac{1}{(a+b)}~~~,~~~~~~~m_2=\dd\frac{1}{a}~~~,~~~~~~~m_3=\dd\frac{1}{(b-a)}
\label{numasses}
\ee
and $U$ is given by
\be
U=\left(
\begin{array}{ccc}
\sqrt{2/3}& 1/\sqrt{3}& 0\\
-1/\sqrt{6}& 1/\sqrt{3}& -1/\sqrt{2}\\
-1/\sqrt{6}& 1/\sqrt{3}& +1/\sqrt{2}
\end{array}
\right)~~~.
\label{HPSmatrix1}
\ee
Note that, in the leading approximation, the model predicts the relation:
\be
\frac{2}{m_2}=\frac{1}{m_1}-\frac{1}{m_3}~~~.
\ee
 It is expected to hold up to corrections
of $O(\lambda^2)$, as will be discussed in section 5. Notice, 
that in our conventions $m_i$ $(i=1,2,3)$
are in general complex numbers, so that the previous relation cannot
be used to exactly predict one physical neutrino mass in terms of the other 
two ones. Nevertheless, it provides a non-trivial constraint that the neutrino 
masses should obey.
 
To get the right solar mixing angle, we should impose $|m_2|>|m_1|$ and this requires $\cos\phi>-|z|/2$, where $z=b/a$ and $\phi$ is the phase difference
between the complex numbers $a$ and $b$. The neutrino spectrum can have 
 either normal or inverted mass ordering. 
If $\max(-1,-|z|/2)\le\cos\phi\le0$
the ordering is inverted, $|m_3|\le |m_1|<|m_2|$, while $|z|/2\le \cos\phi\le 1$ gives rise to a normal ordering, $|m_1|<|m_2|\le |m_3|$.
By defining 
\be
r\equiv \Delta m^2_{sol}/\Delta m^2_{atm}~~~,~~~~~~~
\Delta m^2_{sol}\equiv |m_2|^2-|m_1|^2~~~,~~~~~~~
\Delta m^2_{atm}\equiv\left\vert|m_3|^2-|m_1|^2\right\vert~~~,
\ee
we find
\be
r=\frac{|1-z|^2|z+\bar{z}+|z|^2|}{2|z+\bar{z}|}~~~,~~~~~~~~~~~z\equiv\frac{b}{a}~~~.
\label{rrr}
\ee
We see that a sufficiently small $r$ requires $z$ not to far from either +1 ($\cos\phi=1$, normal hierarchy) or -2 ($\cos\phi=-1$, inverted hierarchy).
If we expand $z$ around +1, we obtain:
\bea
|m_1|^2&=&\frac{1}{3}
\Delta m^2_{atm}~r+...\nn\\
|m_2|^2&=&\frac{4}{3}
\Delta m^2_{atm}~r+...\nn\\
|m_3|^2&=&\left(1+\frac{r}{3}\right)
\Delta m^2_{atm}+...\nn\\
|m_{ee}|^2&=&\frac{16 }{27}\Delta m^2_{atm}~r+...~~~,
\label{lospe}
\eea
where we have expressed the parameters in terms of $\Delta m^2_{atm}$ and $r$. Dots denote terms of order $r^2$ and $|m_{ee}|$ is the effective mass
combination controlling the violation of the total lepton number in neutrinoless double beta decay. 
It is useful to estimate the cut-off $\Lambda$. We have roughly 
\be
\sqrt{\Delta m^2_{atm}}\approx \dd\frac{s^2 (v_u^0)^2}{|a| \Lambda \sqrt{r}}~~~.
\ee
By taking $\sqrt{\Delta m^2_{atm}}=0.05$ eV, $s^2 (v_u^0)^2=(100$ GeV$)^2$ and  $\sqrt{r}\approx 0.2$, we obtain $|a| \Lambda\approx 10^{15}$ GeV, not far from the
unification scale. For $u\approx v_{S,T}\approx \lambda^2$
 the cut-off $\Lambda$ is then above $10^{16}$ GeV.
If we expand $z$ around -2, we get:
\bea
|m_1|^2&=&\left(\dd\frac{9}{8}+\dd\frac{r}{12}\right)
\Delta m^2_{atm}+...\nn\\
|m_2|^2&=&\left(\dd\frac{9}{8}+\dd\frac{13}{12}r\right)
\Delta m^2_{atm}+...\nn\\
|m_3|^2&=&\left(\dd\frac{1}{8}+\dd\frac{r}{12}\right)
\Delta m^2_{atm}+...\nn\\
|m_{ee}|^2&=&\left(\dd\frac{1}{8}-\dd\frac{11}{108}r\right)
\Delta m^2_{atm}+...
~~~.
\label{lospe2}
\eea
We now have
\be
\sqrt{\Delta m^2_{atm}}\approx \dd\frac{s^2 (v_u^0)^2}{|a|\Lambda}~~~.
\ee
By repeating the previous estimate, we find $|a|\Lambda \approx 10^{14}$  GeV and $\Lambda$ slightly below $10^{16}$ GeV.

 Several remarks should be made:

 Concerning the lepton mixing, this is dominated by $U$, eq. (\ref{HPSmatrix1}). The contribution from the charged lepton sector depends
on the entries denoted by the dots in $m_e$. 
Putting all the dots to zero, the charged leptons affect the lepton mixing 
 through rotations of order $\lambda^4$, $\lambda^8$ and $\lambda^4$ in the 12, 13 and 23 sectors, respectively.
Operators of dimensions higher than the ones, considered so far, are strongly 
suppressed and provide contributions of order $\lambda^4$ to 
the mixing matrix. These are negligible, since the leading effect comes from 
the modification of the vacuum structure of eq. (\ref{align}), due to higher
order terms in the scalar potential. We shall discuss this in sections 4 and 5. 
 Eventually, such terms modify only slightly the TB mixing pattern.

 Apart from $w_{\nu}$ contributions to neutrino masses and mixing angles might come from higher dimensional operators, as for instance
\be
\frac{\xi \xi F F H_5 H_5}{\Lambda^4}~~~.
\ee
 However, they are completely negligible compared to those discussed above. If we forced this type of operator to be the dominant one, by eliminating
the singlets $N$ from our model, we would need a value of $\Lambda$ too small compared with the GUT scale.

Depending on the value of $z$, our model gives rise to two separate branches in the neutrino spectrum, both characterized by a nearly
TB mixing. On the first branch, $z\approx+1$, we find a spectrum with normal hierarchy, while on the second branch, $z\approx-2$, we get an inverted hierarchy.
A degenerate spectrum is actually disfavored in our construction, since 
 it would require $z\ll1$ (see eq. (\ref{numasses})) which leads to $r$ close to 1/2,
 as can be read off from eq. (\ref{rrr}). This can obviously not be reconciled with the data. 

In our model the possibility of normal hierarchy is somewhat more natural than the one of inverted hierarchy.
There is no reason a priori why $z$ should be close to $+1$ or to $-2$ and reproducing $r$ requires some amount of tuning.
However, such a tuning is stronger for inverted hierarchy ({\tt ih}) than for the normal one ({\tt nh}), as can be seen by
\be
\left.\frac{d r}{d z}\right\vert_{\tt nh} \left.\frac{d z}{d
  r}\right\vert_{\tt ih}\approx -\frac{4}{3 \sqrt{3}} \sqrt{r}\approx
- 0.14~~~.
\ee
 The derivatives are
computed at the relevant value of $z$ in each case and $r$ is the
experimental value. Moreover the solution with a normal hierarchy has a domain of validity in energy larger by a factor of $1/\sqrt{r}\approx 5.6$
and extends beyond $10^{16}$ GeV.
In the normal hierarchy solution we find with the help of  eq. (\ref{lospe}) 
\be
\sum_i |m_i|\approx (0.06 - 0.07)~{\rm eV}~~~~~~~~~\mbox{and}~~~~~~~~|m_{ee}|\approx 0.007~{\rm eV}~~~~~.
\ee
It is interesting to see that $|m_{ee}|$ is close to the upper limit of the range expected in the normal hierarchy case, being not too far from
the aimed for sensitivity of the next generation of neutrinoless double beta decay experiments, $0.01$ eV.
This is partly attributed to the fact that $|m_1|\approx 0.005$ is different from zero and in part to the absence of a negative interference 
with the $m_3$ contribution, as $\theta_{13}=0$.

%
%%%%%%%%%%%%%%%%%%%%%%%% 4.SECTION: VACUUM ALIGNMENT   %%%%%%%%%%%%%%%%%%%%%%%%%%%%%%
%

\section{Vacuum Alignment}

Here we discuss the minimisation of the scalar potential, in order to justify the VEVs assumed in the previous section. We work in the limit
of exact SUSY. This will not allow us to analyse the electroweak symmetry breaking induced by $H_{5}$ and $H_{\bar 5}$, whose VEVs
are assumed to vanish in first approximation. Indeed all the VEVs we are interested in here, 
 i.e. those of the flavon fields $\varphi_{S,T}$, $\xi$, $\tilde{\xi}$,
$\theta$ and $\theta''$, are relatively close in magnitude to the cut-off $\Lambda$ and therefore much larger than the electroweak scale, which will be consistently
neglected. 
Moreover we work at leading order in the parameter $1/\Lambda$, that is we keep only the lowest dimensional operators in the superpotential
shown in the previous section. Subleading effects will be discussed
 later on. All the multiplets but the flavon ones are assumed to have vanishing 
VEVs and set to zero for the present discussion. 
We regard the U(1) Froggatt-Nielsen flavour symmetry as local. Since the field content displayed in table 1 is anomalous under the U(1),
we need additional chiral multiplets to cancel the anomaly. These multiplets can be chosen vector-like with respect to SU(5), so that they only contribute
to the U(1) anomaly. Here we do not need to specify these fields, but we must 
 presume that they do not acquire a VEV.
Within these assumptions the relevant part of the scalar potential of the model is given by the sum of 
 the F-terms and of a D-term:
\be
V=V_F+V_D~~~,
\ee
\be
V_F=\sum_i\left\vert\frac{\partial w}{\partial\varphi_i}\right\vert^2~~~,
\ee
where $\varphi_i$ stands for the generic chiral multiplet. Only the last term in eq. (\ref{superw}), $w_d$, contributes to the VEVs we are looking for.
It is given by:
\bea
w_d&=&M (\varphi_0^T \varphi_T)+ g (\varphi_0^T \varphi_T\varphi_T)\nn\\
&+&g_1 (\varphi_0^S \varphi_S\varphi_S)+
g_2 \tilde{\xi} (\varphi_0^S \varphi_S)+
g_3 \xi_0 (\varphi_S\varphi_S)+
g_4 \xi_0 \xi^2+
g_5 \xi_0 \xi \tilde{\xi}+
g_6 \xi_0 \tilde{\xi}^2~~~~.
%\label{wd}
\nn
\eea
Since also the terms in $w_d$ have to have $R$-charge +2, we introduce additional
gauge singlets, so called driving fields, $\varphi_{0}^{T}$, $\varphi_{0}^{S}$ and 
$\xi_{0}$ with $R$-charge +2 (see table 1). Note that therefore all terms in $w_d$ are linear
in these fields. Note further that due to U(1) invariance neither the multiplet 
$\theta$, nor the multiplet $\theta''$ is contained in $w_d$.
 Moreover the D-term $V_D$ does not depend on $\varphi_{S,T}$,
  $\xi$, $\tilde{\xi}$, which are all singlets under the (gauged) U(1).
The expression of $w_d$ and the minimisation procedure are exactly as
described in ref. \cite{Mod} and leads to the result anticipated in
the previous section:
\begin{eqnarray}\nonumber
&&\langle\varphi_T\rangle=(v_T,0,0)\Lambda~~~,~~~~v_T\Lambda=-\dd\frac{3M}{2g}~~~,\\ \nonumber
&&\langle\varphi_S\rangle=(v_S,v_S,v_S)\Lambda~~~,~~~~v_S=\dd\frac{\tilde{g}_4}{3 \tilde{g}_3} u~~~,\\ \nonumber
&&\langle\xi\rangle=u\Lambda~~~,\\
&&\langle\tilde{\xi}\rangle=0
\label{sols}
\end{eqnarray}
with $u$ undetermined and $g_3\equiv 3\tilde{g}_3^2$ , $g_4\equiv -\tilde{g}_4^2$. In the following
 we take $v_T$, $v_S$ and $u$ to be of $O(\lambda^2)$. This order of magnitude is indicated by the observed ratio of up and down or charged lepton masses, by the scale of the light neutrino masses and is 
 also compatible with the bounds on the deviations from TB mixing for leptons.

The D-term is given by:\footnote{Note that $\vert\theta''\vert^2$ is a singlet under $A_4$,
 because $\theta''\sim1''$ and $\theta''^*\sim1'$ under $A_4$.}
\be
V_D=\frac{1}{2}(M_{FI}^2- g_{FN}\vert\theta\vert^2-g_{FN}\vert\theta''\vert^2+...)^2
\ee
where $g_{FN}$ is the gauge coupling constant of U(1) and $M_{FI}^2$ denotes the contribution of the Fayet-Iliopoulos term. We have omitted the SU(5) contribution to the D-term, whose VEV is 
 zero.
There are SUSY minima such that $V_F=V_D=0$. The vanishing of $V_D$
requires
\be
g_{FN}\vert\theta\vert^2+g_{FN}\vert\theta''\vert^2=M_{FI}^2~~~.
\ee
If the parameter $M_{FI}^2$ is positive, the above condition
determines a non-vanishing VEV for a combination of $\theta$ and
$\theta''$. Here we assume that the VEVs fulfil $t$, $t'' \sim O(\lambda)$ 
 according to eqs. (\ref{align},\ref{ttpr_order}). The different order of $t$, $t''$ versus $v_T$, $v_S$ and $u$ 
can be attributed to the different couplings and mass parameters in $V_D$ and $V_F$.

Finally, we discuss the subleading corrections to the vacuum alignment.  As already noticed
above, the fields $\theta$ and ${\theta}''$ cannot couple to the flavon fields, since the flavons $\varphi_T$,
$\varphi_S$, $\xi$, $\tilde{\xi}$, $\varphi ^T _0$, $\varphi ^S _0$ and $\xi _0$ are not charged under the U(1)
symmetry, responsible for the charged fermion mass hierarchy. Therefore, the subleading effects in the potential
 arise from terms made up of one driving field and three fields $\varphi_T$, $\varphi_S$, $\xi$ and $\tilde{\xi}$. 
They induce shifts in the VEVs shown above
and thereby influence the mass matrices, as discussed in the next section. Since the flavon field content of this model is essentially the
same as the one in ref. \cite{Mod}, not only the renormalizable part of $w_d$ coincides, but also the
subleading terms are the same. Hence, we do not need to repeat this discussion and we only state the results found there. The shifted VEVs are
\begin{eqnarray}\nonumber
\langle\varphi_T\rangle/\Lambda&=&(v_T+\delta v_{T1},\delta v_{T2},\delta v_{T3})~~~,\\ \nonumber
\langle\varphi_S\rangle/\Lambda&=&(v_S+\delta v_1,v_S+\delta v_2,v_S+\delta v_3)~~~,\\ \nonumber
\langle\xi\rangle/\Lambda&=&u~~~,\\ \label{sols1}
\langle\tilde{\xi}\rangle/\Lambda&=&\delta u'~~~,
\end{eqnarray}
where $u$ remains undetermined and, once we have taken $v_{T,S}$,  $u \sim O(\lambda^2)$, all shifts are suppressed by a factor of order $\lambda^2$: $\delta v/v \sim O(\lambda^2)$.
 As found in ref. \cite{Mod} the following relation holds:
\be
\delta v_{T2} = \delta v_{T3}~~~.\\
\label{deltas}
\ee
Higher order corrections to $t$ and $t^{\prime\prime}$ simply amount to a rescaling that does not change their individual order of magnitude which remains of $O(\lambda)$.

%%%%%%%%%%%%%%%%%%%%%%%% 5.  SUBLEADING CORRECTIONS   %%%%%%%%%%%%%%%%%%%%%%%%%%%%%%

\section{Subleading Corrections}

In this section, we analyse the effects of the subleading corrections in terms of $\lambda$ to the fermion masses and mixings. The corrections arise from additional insertions of the flavons $\varphi_T$,
$\varphi_S$, $\xi$ and $\tilde{\xi}$ as well as from shifts of the VEVs shown above.

%%%%%%%%%%%%%%%%%%%%%%%% 5.1.  CORRECTIONS TO W_UP   %%%%%%%%%%%%%%%%%%%%%%%%%%%%%%

\mathversion{bold}
\subsection{Corrections to $w_{up}$}
\mathversion{normal}

In the up quark sector the leading order terms only involve the fields $\theta$ and $\theta''$, since they are
the only fields which have a non-vanishing U(1) charge among the gauge singlets of the model. The subleading
terms then additionally involve the fields $\varphi_{T}$, $\varphi_{S}$, $\xi$ 
and $\tilde{\xi}$. As the tenplets transform as singlets under $A_4$ and the combinations $T_i T_j H_5 \theta^n {\theta''}^m$ 
are invariant under the $Z_3$ group, we cannot
multiply the $w_{up}$ terms by a single flavon field. The most economic possibility is to insert two flavons, namely $\varphi_{T}
\, \varphi_{T}$. Among the three contractions leading to a 1 or $1'$ or $1''$ representation of $A_4$ only the 1 has a non-vanishing VEV, given that $\langle \varphi_{T} \rangle = (v_{T},0,0) \, \Lambda$. Therefore the dominant subleading
corrections to the up quark mass matrix have the same structure as the
leading order results and are suppressed by an
overall factor $v_{T}^2 \sim O(\lambda^4)$. The fields $\varphi_{S}$ and $\xi$, $\tilde{\xi}$ can only couple
at the level of three flavon insertions due to the requirement of $Z_3$ invariance. However, all contributions
stemming from three flavon insertions are suppressed by $\lambda^6$ relative to the leading order term. Similarly,
the corrections due to shifts in the VEVs contribute at most at relative order $\lambda^6$.
 For the up quark masses and the mixings all these corrections are negligible.

%%%%%%%%%%%%%%%%%%%%%%%% 5.2.  CORRECTIONS TO W_DOWN   %%%%%%%%%%%%%%%%%%%%%%%%%%%%%%

\mathversion{bold}
\subsection{Corrections to $w_{down}$}
\mathversion{normal}

In the down sector the main effect of the subleading corrections is to fill the zeros indicated
by dots in the upper triangle of $m_d$. In order to maintain the $A_4$ invariance the leading order terms include one
insertion of the flavon $\varphi_T$. The subleading corrections arise from two effects: $a.)$ replacing  $\varphi_T$ with products of flavon fields and $b.)$ including the corrections to the VEVs of $\varphi_T$. The replacement of $\varphi_T$ with
a product $\varphi_T \varphi_T$ is the simplest choice compatible with the $Z_3$ charges.
 Note that this is similar to the up quark sector. If the VEVs are unchanged this contribution to $m_d$
is of the same form as displayed in eq. (\ref{md}) and suppressed by $v_T \sim O(\lambda^2)$ compared to the 
leading result due
to the additional flavon field. Therefore this type of correction does not fill the zeros in $m_d$.
They are filled by the corrections coming from the VEV shifts inserted in the terms containing one flavon $\varphi_{T}$.
 Considering that we assumed all $\delta v/v \sim O(\lambda^2)$, the corrections to the matrix elements 
of $m_d$ are of the following order in $\lambda$:
\begin{equation}
\nonumber
\delta m_d=\left(
\begin{array}{ccc}
\lambda^6 & \lambda^4 & \lambda^2 \\
\lambda^6 & \lambda^4 & \lambda^2 \\
\lambda^6 & \lambda^4 & \lambda^2
\end{array}
\right) v_T \lambda v^0_d ~~~.
\label{corrvev1}
\end{equation}
As said, the matrix elements which are already non-vanishing at the leading order, eq. (\ref{md}), 
receive additional corrections from the two flavon insertion $\varphi_T \varphi_T$. These are of the 
same order 
as the corrections from the VEV shifts, e.g. for the element $11$ also of order $\lambda^6$.
In summary, the zeroes in the elements $12$, $13$ and
$23$ of $m_d$, appearing at leading order, are replaced by terms of order $\lambda^4$, $\lambda^2$
and $\lambda^2$, respectively, in units of $ v_T \lambda v^0_d$. 

In our model the relation $m_d=m_e^T$ is not valid for the first two families but it still holds at the level of orders of magnitude for each entry. So the powers of $\lambda$ are also the same for each matrix element of $m_d  m_d^\dagger$ and of $m_e^\dagger  m_e$. This is important as the matrix $m_e^\dagger  m_e$ is diagonalized by the unitary matrix $U_e$ that enters in determining the leptonic mixing matrix $U=U_e^\dagger U_{\nu}$. The results just described for the subleading corrections on $m_d$ and $m_e^\dagger$ imply that $U_e$ induces corrections of $O(\lambda^2)$ on all mixing angles in $U$, that is, in our case, corrections of $O(\lambda^2)$ to the TB values of each mixing angle.

%%%%%%%%%%%%%%%%%%%%%%%% 5.3.  CORRECTIONS TO W_NU   %%%%%%%%%%%%%%%%%%%%%%%%%%%%%%

\mathversion{bold}
\subsection{Corrections to $w_{\nu}$}
\mathversion{normal}

Also the $w_{\nu}$ term of the superpotential, eq. (\ref{wnu}), is
modified by terms with more flavon factors and by subleading
corrections to the VEVs. The Dirac mass term, proportional to $H_5(NF)$,
is mainly modified by a single $\varphi_T$ insertion,
that produces corrective terms suppressed by a $O(\lambda^2)$
  factor. These corrections are of the same order as those arising for Majorana mass terms.
 In fact, $NN$ can be in a $1$ , $1'$, $1''$ or $3_s$ combination. Since $NN\sim \omega^2$ under $Z_3$, 
the singlet $1$ can be multiplied by $\xi$ (the singlet leading term) or by $(\varphi_T \varphi_S)$ 
(which can be absorbed into a redefinition of the leading term), $1'$ by $(\varphi_T \varphi_S)^{''}$, $1''$ 
by $(\varphi_T \varphi_S)^{\prime}$ and $3_s$ by $ \varphi_S$ (the triplet leading term) or 
by $(\varphi_T \xi)$ or $(\varphi_T \varphi_S)_{3_s}$ or $(\varphi_T \varphi_S)_{3_a}$. All 
 two flavon insertions lead to corrections of relative order of $O(\lambda^2)$ to the 
matrix elements of the Majorana matrix. In addition, the shifts of the $\varphi_S$ VEVs applied to the triplet leading term also produce $O(\lambda^2)$ corrective terms. As it is easy to check, in general there are enough parameters so that all 6 independent entries of the (symmetric) Majorana mass matrix receive a different correction at $O(\lambda^2)$.

 The described corrections affect the neutrino masses and, together with the corrections to $m_e$, also 
 all lepton mixing angles. To be compatible with the data, given the accuracy of the TB approximation, 
the dominant corrections must be of $O(\lambda^2)$ at most, and this is precisely the magnitude of the 
terms that we have just mentioned.

%
%%%%%%%%%%%%%%%%%%%%%%%% 6. SECTION: CONCLUSIONS   %%%%%%%%%%%%%%%%%%%%%%%%%%%%%%
%
\section{Conclusion}

We have constructed a SUSY SU(5) grand unified model which includes
the $A_4$ description of TB mixing for leptons. For this it is not only
necessary to adopt an $A_4$ classification of quarks and leptons
compatible with SU(5), but also to introduce additional U(1) and $Z_N$
symmetries and to suitably formulate the grand unification model. We
find that the most attractive solution to cope with the different
requirements from fermion mass and mixing hierarchies, from the
problem of doublet-triplet splitting in the Higgs sector, from proton
decay bounds and from maintaining bottom tau unification only, 
 is a formulation in 5 space-time dimensions with a
particular location of the different fields, with some of them on the
brane at $y=0$ and some in the bulk. The latter include the gauge
and Higgs fields as well as the tenplets of the first two, i.e. lightest,
families. The resulting model naturally leads to TB mixing in first
approximation with corrections of $O(\lambda^2)$ from higher dimensional
effective operators, together with reproducing the observed mass
hierarchies for quarks and charged leptons and the CKM mixing pattern.
 In the quark sector, however, as is typical of U(1) models, only orders of
magnitude are determined in terms of powers of $\lambda$ with
exponents fixed by the charges. A moderate fine tuning is only needed
to enhance the CKM mixing angle between the first two generations,
which would generically be of $O(\lambda^2)$, and to suppress
the value of $r$, given in eq. (\ref{rrr}), which would typically be
of order 1. The latter feature is also true in all purely leptonic $A_4$
models, in which $A_4$ leads to the correct mixing, but not directly to the
spectrum of the neutrino masses. Actually the model allows for both
types of neutrino mass hierarchy, the normal and the inverted one.
The normal hierarchy is, however, somewhat more natural, since it
requires less tuning to reproduce $r$. Furthermore, it is consistent with a
larger value of the cut-off $\Lambda$. If the normal
hierarchy is the correct one, the model predicts the sum of
neutrino masses to be around $(0.06-0.07)$ eV and $|m_{ee}|$ 
 to be close to 0.007 eV.
 Therefore, $|m_{ee}|$ is not far from the future aimed for 
experimental sensitivity. Finally, all subleading corrections to the leading order result 
of fermion masses and mixings have been carefully analysed.
 
In conclusion, we have demonstrated that the simple $A_4$ approach to TB mixing is compatible with a grand unified picture describing all quark and lepton masses and mixings.

%%%%%%%%%%%%%%%%%%%%%%%% ACKNOWLEDGEMENTS   %%%%%%%%%%%%%%%%%%%%%%%%%%%%%%

\section*{Acknowledgements}

We thank S. King for very interesting discussions. We recognize that this work has been partly supported by the European Commission under contracts MRTN-CT-2004-503369 and MRTN-CT-2006-035505, and by the Italian Ministero dell'Universita' e della Ricerca Scientifica, under the COFIN program for 2007-08.

\vfill

%%%%%%%%%%%%%%%%%%%%%%%% REFERENCES   %%%%%%%%%%%%%%%%%%%%%%%%%%%%%%


\begin{thebibliography}{99}

\bibitem{data}
T.~Schwetz,
  %``Neutrino oscillations: Current status and prospects,''
  Acta Phys.\ Polon.\  B {\bf 36} (2005) 3203
  [arXiv:hep-ph/0510331];
 %%CITATION = APPOA,B36,3203;%%
A.~Strumia and F.~Vissani,
  %``Implications of neutrino data circa 2005,''
  Nucl.\ Phys.\  B {\bf 726} (2005) 294
  [arXiv:hep-ph/0503246];
  %%CITATION = NUPHA,B726,294;%%
G.~L.~Fogli, E.~Lisi, A.~Marrone, A.~Melchiorri, A.~Palazzo, P.~Serra and J.~Silk,
  %``Observables sensitive to absolute neutrino masses: Constraints and
  %correlations from world neutrino data,''
  Phys.\ Rev.\ D {\bf 70} (2004) 113003
  [arXiv:hep-ph/0408045];
  %%CITATION = HEP-PH 0408045;%%
J.~N.~Bahcall, M.~C.~Gonzalez-Garcia and C.~Pena-Garay,
  %``Solar neutrinos before and after Neutrino 2004,''
  JHEP {\bf 0408}, 016 (2004)
  [arXiv:hep-ph/0406294];
  %%CITATION = HEP-PH 0406294;%%
M.~Maltoni, T.~Schwetz, M.~A.~Tortola and J.~W.~F.~Valle,
  %``Status of global fits to neutrino oscillations,''
  New J.\ Phys.\  {\bf 6} (2004) 122
  [arXiv:hep-ph/0405172].
  %%CITATION = HEP-PH 0405172;%%
\bibitem{hps}
P.~F.~Harrison, D.~H.~Perkins and W.~G.~Scott,
  %``Tri-bimaximal mixing and the neutrino oscillation data,''
  Phys.\ Lett.\ B {\bf 530} (2002) 167
  [arXiv:hep-ph/0202074];
  %%CITATION = HEP-PH 0202074;%%
P.~F.~Harrison and W.~G.~Scott,
  %``Symmetries and generalisations of tri-bimaximal neutrino mixing,''
  Phys.\ Lett.\ B {\bf 535} (2002) 163
  [arXiv:hep-ph/0203209];
  %%CITATION = HEP-PH 0203209;%%
Z.~z.~Xing,
  %``Nearly tri-bimaximal neutrino mixing and CP violation,''
  Phys.\ Lett.\ B {\bf 533} (2002) 85
  [arXiv:hep-ph/0204049];
  %%CITATION = HEP-PH 0204049;%%
P.~F.~Harrison and W.~G.~Scott,
  %``mu - tau reflection symmetry in lepton mixing and neutrino oscillations,''
  Phys.\ Lett.\ B {\bf 547} (2002) 219
  [arXiv:hep-ph/0210197];
  %%CITATION = HEP-PH 0210197;%%
P.~F.~Harrison and W.~G.~Scott,
  %``Permutation symmetry, tri-bimaximal neutrino mixing and the S3 group
  %characters,''
  Phys.\ Lett.\ B {\bf 557} (2003) 76
  [arXiv:hep-ph/0302025];
  %%CITATION = HEP-PH 0302025;%%
P.~F.~Harrison and W.~G.~Scott,
  %``Status of tri- / bi-maximal neutrino mixing,''
  arXiv:hep-ph/0402006;
  %%CITATION = HEP-PH 0402006;%%
P.~F.~Harrison and W.~G.~Scott,
  %``The simplest neutrino mass matrix,''
  Phys.\ Lett.\  B {\bf 594} (2004) 324
  [arXiv:hep-ph/0403278].
  %%CITATION = PHLTA,B594,324;%%

\bibitem{ma1}
E.~Ma and G.~Rajasekaran,
  %``Softly broken A(4) symmetry for nearly degenerate neutrino masses,''
  Phys.\ Rev.\ D {\bf 64} (2001) 113012
  [arXiv:hep-ph/0106291].
  %%CITATION = HEP-PH 0106291;%
  
 \bibitem{ma2}
K.~S.~Babu, E.~Ma and J.~W.~F.~Valle,
  %``Underlying A(4) symmetry for the neutrino mass matrix and the quark  mixing
  %matrix,''
  Phys.\ Lett.\ B {\bf 552} (2003) 207
  [arXiv:hep-ph/0206292];
  %%CITATION = HEP-PH 0206292;%%
M.~Hirsch, J.~C.~Romao, S.~Skadhauge, J.~W.~F.~Valle and A.~Villanova del Moral,
  %``Degenerate neutrinos from a supersymmetric A(4) model,''
  arXiv:hep-ph/0312244;
  %%CITATION = HEP-PH 0312244;%%
M.~Hirsch, J.~C.~Romao, S.~Skadhauge, J.~W.~F.~Valle and A.~Villanova del Moral,
  %``Phenomenological tests of supersymmetric A(4) family symmetry model of
  %neutrino mass,''
  Phys.\ Rev.\  D {\bf 69} (2004) 093006
  [arXiv:hep-ph/0312265];
  %%CITATION = PHRVA,D69,093006;%%
E.~Ma,
  %``A(4) origin of the neutrino mass matrix,''
  Phys.\ Rev.\  D {\bf 70} (2004) 031901
  [arXiv:hep-ph/0404199];
  %%CITATION = PHRVA,D70,031901;%%
E.~Ma
%``Non-Abelian discrete family symmetries of leptons and quarks,''
  arXiv:hep-ph/0409075;
  %%CITATION = HEP-PH 0409075;%%
E.~Ma,
  %``Symmetries and neutrino masses,''
  New J.\ Phys.\  {\bf 6} (2004) 104
  [arXiv:hep-ph/0405152];
  %%CITATION = NJOPF,6,104;%%
S.~L.~Chen, M.~Frigerio and E.~Ma,
  %``Hybrid seesaw neutrino masses with A(4) family symmetry,''
  Nucl.\ Phys.\  B {\bf 724} (2005) 423
  [arXiv:hep-ph/0504181];
  %%CITATION = NUPHA,B724,423;%%
E.~Ma,
  %``Aspects of the tetrahedral neutrino mass matrix,''
  Phys.\ Rev.\ D {\bf 72} (2005) 037301
  [arXiv:hep-ph/0505209];
  %%CITATION = HEP-PH 0505209;%%
K.~S.~Babu and X.~G.~He,
  %``Model of geometric neutrino mixing,''
  arXiv:hep-ph/0507217;
  %%CITATION = HEP-PH 0507217;%%
A.~Zee,
  %``Obtaining the neutrino mixing matrix with the tetrahedral group,''
  Phys.\ Lett.\ B {\bf 630} (2005) 58
  [arXiv:hep-ph/0508278];
  %%CITATION = HEP-PH 0508278;%%
E.~Ma,
  %``Tetrahedral family symmetry and the neutrino mixing matrix,''
  Mod.\ Phys.\ Lett.\ A {\bf 20} (2005) 2601
  [arXiv:hep-ph/0508099];
  %%CITATION = HEP-PH 0508099;%%
E.~Ma,
  %``Tribimaximal neutrino mixing from a supersymmetric model with A4 family
  %symmetry,''
  Phys.\ Rev.\  D {\bf 73} (2006) 057304
  [arXiv:hep-ph/0511133];
  %%CITATION = PHRVA,D73,057304;%%
S.~K.~Kang, Z.~z.~Xing and S.~Zhou,
  %``Possible deviation from the tri-bimaximal neutrino mixing in a seesaw
  %model,''
  Phys.\ Rev.\  D {\bf 73} (2006) 013001
  [arXiv:hep-ph/0511157];
  %%CITATION = PHRVA,D73,013001;%%
X.~G.~He, Y.~Y.~Keum and R.~R.~Volkas,
  %``A(4) flavour symmetry breaking scheme for understanding quark and  neutrino
  %mixing angles,''
  JHEP {\bf 0604} (2006) 039
  [arXiv:hep-ph/0601001].
  %%CITATION = JHEPA,0604,039;%%

\bibitem{OurTriBi}
G.~Altarelli and F.~Feruglio,
  %``Tri-bimaximal neutrino mixing from discrete symmetry in extra dimensions,''
  Nucl.\ Phys.\ B {\bf 720} (2005) 64
  [arXiv:hep-ph/0504165].
  %%CITATION = HEP-PH 0504165;%%
 
 \bibitem{Mod}
G.~Altarelli and F.~Feruglio,
  %``Tri-bimaximal neutrino mixing, A(4) and the modular symmetry,''
  Nucl.\ Phys.\ B {\bf 741} (2006) 215
  [arXiv:hep-ph/0512103].
  
\bibitem{us3} 
G.~Altarelli, F.~Feruglio and Y.~Lin,
  %``Tri-bimaximal neutrino mixing from orbifolding,''
  Nucl.\ Phys.\  B {\bf 775} (2007) 31
  [arXiv:hep-ph/0610165].
  %%CITATION = NUPHA,B775,31;%%
 
\bibitem{continuous}
S.~F.~King,
  %``Predicting neutrino parameters from SO(3) family symmetry and quark-lepton
  %unification,''
  JHEP {\bf 0508} (2005) 105
  [arXiv:hep-ph/0506297];
  %%CITATION = HEP-PH 0506297;%%
I.~de Medeiros Varzielas and G.~G.~Ross,
  %``SU(3) family symmetry and neutrino bi-tri-maximal mixing,''
  Nucl.\ Phys.\ B {\bf 733} (2006) 31
  [arXiv:hep-ph/0507176];
  %%CITATION = HEP-PH 0507176;%%
S.~F.~King and M.~Malinsky,
  %``Towards a complete theory of fermion masses and mixings with SO(3)  family
  %symmetry and 5d SO(10) unification,''
  JHEP {\bf 0611} (2006) 071
  [arXiv:hep-ph/0608021];
  %%CITATION = JHEPA,0611,071;%%
Y.~Koide,
  %``Broken SU(3) Flavor Symmetry and Tribimaximal Neutrino Mixing,''
  arXiv:0707.0899 [hep-ph].
  %%CITATION = ARXIV:0707.0899;%%

\bibitem{others}
C.~I.~Low and R.~R.~Volkas,
  %``Tri-bimaximal mixing, discrete family symmetries, and a conjecture
  %connecting the quark and lepton mixing matrices,''
  Phys.\ Rev.\  D {\bf 68} (2003) 033007
  [arXiv:hep-ph/0305243];
  %%CITATION = PHRVA,D68,033007;%%
J.~Matias and C.~P.~Burgess,
  %``MSLED, neutrino oscillations and the cosmological constant,''
  JHEP {\bf 0509} (2005) 052
  [arXiv:hep-ph/0508156];
  %%CITATION = HEP-PH 0508156;%%
E.~Ma,
  %``Neutrino mass matrix from S(4) symmetry,''
  Phys.\ Lett.\  B {\bf 632} (2006) 352
  [arXiv:hep-ph/0508231];
  %%CITATION = PHLTA,B632,352;%%
S.~Luo and Z.~z.~Xing,
  %``Generalized tri-bimaximal neutrino mixing and its sensitivity to  radiative
  %corrections,''
  Phys.\ Lett.\  B {\bf 632} (2006) 341
  [arXiv:hep-ph/0509065];
  %%CITATION = PHLTA,B632,341;%%
I.~de Medeiros Varzielas, S.~F.~King and G.~G.~Ross,
  %``Tri-bimaximal neutrino mixing from discrete subgroups of SU(3) and  SO(3)
  %family symmetry,''
  Phys.\ Lett.\  B {\bf 644} (2007) 153
  [arXiv:hep-ph/0512313];
  %%CITATION = PHLTA,B644,153;%%
N.~Haba, A.~Watanabe and K.~Yoshioka,
  %``Twisted flavors and tri/bi-maximal neutrino mixing,''
  Phys.\ Rev.\ Lett.\  {\bf 97} (2006) 041601
  [arXiv:hep-ph/0603116];
  %%CITATION = PRLTA,97,041601;%%
P.~Kovtun and A.~Zee,
  %``A schematic model of neutrinos,''
  Phys.\ Lett.\ B {\bf 640} (2006) 37
  [arXiv:hep-ph/0604169];
  %%CITATION = HEP-PH 0604169;%%
Z.~z.~Xing, H.~Zhang and S.~Zhou,
  % ``Nearly tri-bimaximal neutrino mixing and CP violation from mu - tau
  %symmetry breaking,''
  Phys.\ Lett.\ B {\bf 641} (2006) 189
  [arXiv:hep-ph/0607091];
  %%CITATION = HEP-PH 0607091;%%
C.~S.~Lam,
  %``Mass independent textures and symmetry,''
  Phys.\ Rev.\  D {\bf 74} (2006) 113004
  [arXiv:hep-ph/0611017];
  %%CITATION = PHRVA,D74,113004;%%
C.~Luhn, S.~Nasri and P.~Ramond,
  %``Tri-Bimaximal Neutrino Mixing and the Family Symmetry Z_7 \rtimes Z_3,''
  Phys.\ Lett.\  B {\bf 652} (2007) 27
  [arXiv:0706.2341 [hep-ph]];
  %%CITATION = PHLTA,B652,27;%%
C.~S.~Lam,
  %``Symmetry of Lepton Mixing,''
  Phys.\ Lett.\  B {\bf 656} (2007) 193
  [arXiv:0708.3665 [hep-ph]];
  %%CITATION = PHLTA,B656,193;%%
E.~Ma,
  %``Near Tribimaximal Neutrino Mixing with Delta(27) Symmetry,''
  arXiv:0709.0507 [hep-ph];
  %%CITATION = ARXIV:0709.0507;%%
E.~Ma,
  %``New lepton family symmetry and neutrino tribimaximal mixing,''
  Europhys.\ Lett.\  {\bf 79} (2007) 61001
  [arXiv:hep-ph/0701016];
  %%CITATION = EULEE,79,61001;%%
C.~S.~Lam,
  %``Horizontal Symmetry,''
  arXiv:0711.3795 [hep-ph].
  %%CITATION = ARXIV:0711.3795;%%

\bibitem{ma1.5}
E.~Ma,
  %``Quark mass matrices in the A(4) model,''
  Mod.\ Phys.\ Lett.\ A {\bf 17} (2002) 627
  [arXiv:hep-ph/0203238].
  %%CITATION = HEP-PH 0203238;%%

\bibitem{maGUT}  
E.~Ma,
  %``Hiding the existence of a family symmetry in the standard model,''
  Mod.\ Phys.\ Lett.\  A {\bf 20} (2005) 2767
  [arXiv:hep-ph/0506036];
  %%CITATION = MPLAE,A20,2767;%%
E.~Ma,
  %``Suitability of A(4) as a family symmetry in grand unification,''
  Mod.\ Phys.\ Lett.\  A {\bf 21} (2006) 2931
  [arXiv:hep-ph/0607190];
  %%CITATION = MPLAE,A21,2931;%%
E.~Ma, H.~Sawanaka and M.~Tanimoto,
  %``Quark masses and mixing with A(4) family symmetry,''
  Phys.\ Lett.\  B {\bf 641} (2006) 301
  [arXiv:hep-ph/0606103];
  %%CITATION = PHLTA,B641,301;%%
S.~Morisi, M.~Picariello and E.~Torrente-Lujan,
  %``A model for fermion masses and lepton mixing in SO(10) x A4,''
  Phys.\ Rev.\  D {\bf 75} (2007) 075015
  [arXiv:hep-ph/0702034];
  %%CITATION = PHRVA,D75,075015;%%
W.~Grimus and H.~Kuhbock,
  %``Embedding the Zee-Wolfenstein neutrino mass matrix in an SO(10) x A4 GUT
  %scenario,''
  arXiv:0710.1585 [hep-ph].
  %%CITATION = ARXIV:0710.1585;%%

\bibitem{KM} 
S.~F.~King and M.~Malinsky,
  %``A(4) family symmetry and quark-lepton unification,''
  Phys.\ Lett.\  B {\bf 645} (2007) 351
  [arXiv:hep-ph/0610250].
  %%CITATION = PHLTA,B645,351;%%

\bibitem{T'0} 
P.~H.~Frampton and T.~W.~Kephart,
  %``Simple nonAbelian finite flavor groups and fermion masses,''
  Int.\ J.\ Mod.\ Phys.\  A {\bf 10} (1995) 4689
  [arXiv:hep-ph/9409330];
  %%CITATION = IMPAE,A10,4689;%%
A.~Aranda, C.~D.~Carone and R.~F.~Lebed,
  %``U(2) flavor physics without U(2) symmetry,''
  Phys.\ Lett.\  B {\bf 474} (2000) 170
  [arXiv:hep-ph/9910392];
  %%CITATION = PHLTA,B474,170;%%
A.~Aranda, C.~D.~Carone and R.~F.~Lebed,
  %``Maximal neutrino mixing from a minimal flavor symmetry,''
  Phys.\ Rev.\  D {\bf 62} (2000) 016009
  [arXiv:hep-ph/0002044];
  %%CITATION = PHRVA,D62,016009;%%
P.~D.~Carr and P.~H.~Frampton,
  %``Group theoretic bases for tribimaximal mixing,''
  arXiv:hep-ph/0701034;
  %%CITATION = HEP-PH/0701034;%%
P.~H.~Frampton and T.~W.~Kephart,
  %``Flavor Symmetry for Quarks and Leptons,''
  JHEP {\bf 0709} (2007) 110
  [arXiv:0706.1186 [hep-ph]];
  %%CITATION = JHEPA,0709,110;%%
A.~Aranda,
  %``Neutrino mixing from the double tetrahedral group $T^{\prime}$,''
  Phys.\ Rev.\  D {\bf 76} (2007) 111301
  [arXiv:0707.3661 [hep-ph]].
  %%CITATION = PHRVA,D76,111301;%%

\bibitem{T'} 
F.~Feruglio, C.~Hagedorn, Y.~Lin and L.~Merlo,
  %``Tri-bimaximal neutrino mixing and quark masses from a discrete flavour
  %symmetry,''
  Nucl.\ Phys.\  B {\bf 775} (2007) 120
  [arXiv:hep-ph/0702194].
  %%CITATION = NUPHA,B775,120;%%

\bibitem{CM} 
M.~C.~Chen and K.~T.~Mahanthappa,
  %``CKM and Tri-bimaximal MNS Matrices in a $SU(5) \times ^{(d)}T$ Model,''
  Phys.\ Lett.\  B {\bf 652} (2007) 34
  [arXiv:0705.0714 [hep-ph]].
  %%CITATION = PHLTA,B652,34;%%
  
\bibitem{S4} 
D.~G.~Lee and R.~N.~Mohapatra,
  %``An SO(10) x S(4) scenario for naturally degenerate neutrinos,''
  Phys.\ Lett.\  B {\bf 329} (1994) 463
  [arXiv:hep-ph/9403201];
  %%CITATION = PHLTA,B329,463;%%
R.~N.~Mohapatra, M.~K.~Parida and G.~Rajasekaran,
  %``High scale mixing unification and large neutrino mixing angles,''
  Phys.\ Rev.\  D {\bf 69} (2004) 053007
  [arXiv:hep-ph/0301234];
  %%CITATION = PHRVA,D69,053007;%%
C.~Hagedorn, M.~Lindner and R.~N.~Mohapatra,
  %``S(4) flavor symmetry and fermion masses: Towards a grand unified theory  of
  %flavor,''
  JHEP {\bf 0606} (2006) 042
  [arXiv:hep-ph/0602244];
  %%CITATION = JHEPA,0606,042;%%
Y.~Cai and H.~B.~Yu,
  %``A SO(10) GUT model with S4 flavor symmetry,''
  Phys.\ Rev.\  D {\bf 74} (2006) 115005
  [arXiv:hep-ph/0608022].
  %%CITATION = PHRVA,D74,115005;%%

\bibitem{27} 
I.~de Medeiros Varzielas, S.~F.~King and G.~G.~Ross,
  %``Neutrino tri-bi-maximal mixing from a non-Abelian discrete family
  %symmetry,''
  Phys.\ Lett.\  B {\bf 648} (2007) 201
  [arXiv:hep-ph/0607045].
  %%CITATION = PHLTA,B648,201;%%

\bibitem{S3} 
R.~Dermisek and S.~Raby,
  %``Bi-large neutrino mixing and CP violation in an SO(10) SUSY GUT for
  %fermion masses,''
  Phys.\ Lett.\  B {\bf 622} (2005) 327
  [arXiv:hep-ph/0507045];
  %%CITATION = PHLTA,B622,327;%%
S.~Morisi and M.~Picariello,
  %``The flavor physics in unified gauge theory from an S(3) x P discrete
  %symmetry,''
  Int.\ J.\ Theor.\ Phys.\  {\bf 45} (2006) 1267
  [arXiv:hep-ph/0505113];
  %%CITATION = IJTPB,45,1267;%%
M.~Picariello,
  %``Neutrino CP violating parameters from nontrivial quark-lepton correlation:
  %A S(3) x GUT model,''
  arXiv:hep-ph/0611189;
  %%CITATION = HEP-PH/0611189;%%
F.~Caravaglios and S.~Morisi,
  %``Fermion masses in E(6) grand unification with family permutation
  %symmetries,''
  arXiv:hep-ph/0510321;
  %%CITATION = HEP-PH 0510321;%%
S.~Morisi,
  %``S(3) family permutation symmetry and quark masses: A model independent
  %approach,''
  arXiv:hep-ph/0604106;
  %%CITATION = HEP-PH/0604106;%%
F.~Caravaglios and S.~Morisi,
  %``Neutrino masses and mixings with an S(3) family permutation symmetry,''
  arXiv:hep-ph/0503234;
  %%CITATION = HEP-PH/0503234;%%
N.~Haba and K.~Yoshioka,
  %``Discrete flavor symmetry, dynamical mass textures, and grand
  %unification,''
  Nucl.\ Phys.\  B {\bf 739} (2006) 254
  [arXiv:hep-ph/0511108];
  %%CITATION = NUPHA,B739,254;%%
M.~Tanimoto and T.~Yanagida,
  %``A higher-dimensional origin of the inverted mass hierarchy for
  %neutrinos,''
  Phys.\ Lett.\  B {\bf 633} (2006) 567
  [arXiv:hep-ph/0511336];
  %%CITATION = PHLTA,B633,567;%%
Y.~Koide,
  %``Seesaw mass matrix model of quarks and leptons with flavor-triplet  Higgs
  %scalars,''
  Eur.\ Phys.\ J.\  C {\bf 48} (2006) 223
  [arXiv:hep-ph/0508301];
  %%CITATION = EPHJA,C48,223;%%
R.~N.~Mohapatra, S.~Nasri and H.~B.~Yu,
  %``Grand unification of mu - tau symmetry,''
  Phys.\ Lett.\  B {\bf 636} (2006) 114
  [arXiv:hep-ph/0603020];
  %%CITATION = PHLTA,B636,114;%%
R.~N.~Mohapatra, S.~Nasri and H.~B.~Yu,
  %``S(3) symmetry and tri-bimaximal mixing,''
  Phys.\ Lett.\  B {\bf 639} (2006) 318
  [arXiv:hep-ph/0605020];
  %%CITATION = PHLTA,B639,318;%%
J.~Kubo, A.~Mondragon, M.~Mondragon and E.~Rodriguez-Jauregui,
  %``The flavor symmetry,''
  Prog.\ Theor.\ Phys.\  {\bf 109} (2003) 795
  [Erratum-ibid.\  {\bf 114} (2005) 287]
  [arXiv:hep-ph/0302196];
  %%CITATION = PTPKA,109,795;%%
J.~Kubo,
  %``Majorana phase in minimal S(3) invariant extension of the standard
  %model,''
  Phys.\ Lett.\  B {\bf 578} (2004) 156
  [Erratum-ibid.\  B {\bf 619} (2005) 387]
  [arXiv:hep-ph/0309167];
  %%CITATION = PHLTA,B578,156;%%
W.~Grimus and L.~Lavoura,
  %``A model realizing the Harrison-Perkins-Scott lepton mixing matrix,''
  JHEP {\bf 0601} (2006) 018
  [arXiv:hep-ph/0509239];
  %%CITATION = JHEPA,0601,018;%%
T.~Teshima,
  %``Flavor mass and mixing and S(3) symmetry: An S(3) invariant model
  %reasonable to all,''
  Phys.\ Rev.\  D {\bf 73} (2006) 045019
  [arXiv:hep-ph/0509094];
  %%CITATION = PHRVA,D73,045019;%%
S.~Kaneko, H.~Sawanaka, T.~Shingai, M.~Tanimoto and K.~Yoshioka,
  %``Flavor symmetry and vacuum aligned mass textures,''
  Prog.\ Theor.\ Phys.\  {\bf 117} (2007) 161
  [arXiv:hep-ph/0609220];
  %%CITATION = PTPKA,117,161;%%
Y.~Koide,
  %``Permutation symmetry S(3) and VEV structure of flavor-triplet Higgs
  %scalars,''
  Phys.\ Rev.\  D {\bf 73} (2006) 057901
  [arXiv:hep-ph/0509214];
  %%CITATION = PHRVA,D73,057901;%%
Y.~Koide,
  %``S(3) symmetry and neutrino masses and mixings,''
  Eur.\ Phys.\ J.\  C {\bf 50} (2007) 809
  [arXiv:hep-ph/0612058];
  %%CITATION = EPHJA,C50,809;%%
C.~Y.~Chen and L.~Wolfenstein,
  %``Consequences of Approximate $S_3$ Symmetry of the Neutrino Mass Matrix,''
  arXiv:0709.3767 [hep-ph];
  %%CITATION = ARXIV:0709.3767;%%
S.~L.~Chen, M.~Frigerio and E.~Ma,
  %``Large neutrino mixing and normal mass hierarchy: A discrete
  %understanding,''
  Phys.\ Rev.\  D {\bf 70} (2004) 073008
  [Erratum-ibid.\  D {\bf 70} (2004) 079905]
  [arXiv:hep-ph/0404084];
  %%CITATION = PHRVA,D70,073008;%%
L.~Lavoura and E.~Ma,
  %``Two predictive supersymmetric S(3) x Z(2) models for the quark mass
  %matrices,''
  Mod.\ Phys.\ Lett.\  A {\bf 20} (2005) 1217
  [arXiv:hep-ph/0502181].
  %%CITATION = MPLAE,A20,1217;%%

\bibitem{5DSU5}
E.~Witten,
  %``Symmetry Breaking Patterns In Superstring Models,''
  Nucl.\ Phys.\  B {\bf 258} (1985) 75;
  %%CITATION = NUPHA,B258,75;%%
Y.~Kawamura,
  %``Triplet-doublet splitting, proton stability and extra dimension,''
  Prog.\ Theor.\ Phys.\  {\bf 105} (2001) 999
  [arXiv:hep-ph/0012125];
  %%CITATION = PTPKA,105,999;%%
A.~E.~Faraggi,
  %``Doublet-triplet splitting in realistic heterotic string derived models,''
  Phys.\ Lett.\  B {\bf 520} (2001) 337
  [arXiv:hep-ph/0107094] and references therein.
  %%CITATION = PHLTA,B520,337;%%

\bibitem{5D}
L.~J.~Hall and Y.~Nomura,
  %``Gauge unification in higher dimensions,''
  Phys.\ Rev.\  D {\bf 64} (2001) 055003
  [arXiv:hep-ph/0103125];
  %%CITATION = PHRVA,D64,055003;%%
Y.~Nomura,
  %``Strongly coupled grand unification in higher dimensions,''
  Phys.\ Rev.\  D {\bf 65} (2002) 085036
  [arXiv:hep-ph/0108170];
  %%CITATION = PHRVA,D65,085036;%%
L.~J.~Hall and Y.~Nomura,
  %``A complete theory of grand unification in five dimensions,''
  Phys.\ Rev.\  D {\bf 66} (2002) 075004
  [arXiv:hep-ph/0205067].
  %%CITATION = PHRVA,D66,075004;%%
  
\bibitem{5Dfreedom}
G.~Altarelli and F.~Feruglio,
  %``SU(5) grand unification in extra dimensions and proton decay,''
  Phys.\ Lett.\  B {\bf 511} (2001) 257
  [arXiv:hep-ph/0102301];
  %%CITATION = PHLTA,B511,257;%%
A.~Hebecker and J.~March-Russell,
  %``A minimal S(1)/(Z(2) x Z'(2)) orbifold GUT,''
  Nucl.\ Phys.\  B {\bf 613} (2001) 3
  [arXiv:hep-ph/0106166];
  %%CITATION = NUPHA,B613,3;%%
A.~Hebecker and J.~March-Russell,
  %``The flavour hierarchy and see-saw neutrinos from bulk masses in 5d
  %orbifold GUTs,''
  Phys.\ Lett.\  B {\bf 541} (2002) 338
  [arXiv:hep-ph/0205143].
  %%CITATION = PHLTA,B541,338;%%

\bibitem{5Dpdecay}
R.~Contino, L.~Pilo, R.~Rattazzi and E.~Trincherini,
  %``Running and matching from 5 to 4 dimensions,''
  Nucl.\ Phys.\  B {\bf 622} (2002) 227
  [arXiv:hep-ph/0108102];
  %%CITATION = NUPHA,B622,227;%%
A.~Hebecker and J.~March-Russell,
  %``Proton decay signatures of orbifold GUTs,''
  Phys.\ Lett.\  B {\bf 539} (2002) 119
  [arXiv:hep-ph/0204037];
  %%CITATION = PHLTA,B539,119;%%
M.~L.~Alciati, F.~Feruglio, Y.~Lin and A.~Varagnolo,
  %``Proton lifetime from SU(5) unification in extra dimensions,''
  JHEP {\bf 0503} (2005) 054
  [arXiv:hep-ph/0501086].
  %%CITATION = JHEPA,0503,054;%%

\bibitem{Nomura}
Y.~Nomura,
  %``Strongly coupled grand unification in higher dimensions,''
  Phys.\ Rev.\  D {\bf 65} (2002) 085036
  [arXiv:hep-ph/0108170].
  %%CITATION = PHRVA,D65,085036;%%
  
\end{thebibliography}
\end{document}